# Experimental demonstration of Single-Level and Multi-Level-Cell RRAM-based In-Memory Computing with up to 16 parallel operations

E. Esmanhotto[1], T. Hirtzlin[1], N. Castellani[1], S. Martin[1], B. Giraud[2], F. Andrieu[1], J.F. Nodin[1], D. Querlioz[4], J-M. Portal[3] and E. Vianello[1]

[1]CEA-Leti, Université Grenoble Alpes, Grenoble, France, email: Eduardo.Esmanhotto@cea.fr, Elisa.Vianello@cea.fr
[2]CEA-List, Université Grenoble Alpes, Grenoble, France, [3]Aix-Marseille Université, IM2NP, Marseille, France,
[4]Université Paris-Saclay, CNRS, C2N, Palaiseau, France.

*Abstract* – Crossbar arrays of resistive memories (RRAM) hold the promise of enabling In-Memory Computing (IMC), but essential challenges due to the impact of device imperfection and device endurance have yet to be overcome. In this work, we demonstrate experimentally an RRAM-based IMC logic concept with strong resilience to RRAM variability, even after one million endurance cycles. Our work relies on a generalization of the concept of in-memory Scouting Logic, and we demonstrate it experimentally with up to 16 parallel devices (operands), a new milestone for RRAM in-memory logic. Moreover, we combine IMC with Multi-Level-Cell programming and demonstrate experimentally, for the first time, an IMC RRAM-based MLC 2-bit adder.

## I. INTRODUCTION

In today's data-intensive applications, data transfer is the main contributor to the power consumption of computing systems, a phenomenon known as the memory wall [1]. To tackle this issue, In-Memory Computing (IMC) [2] is a major lead, especially on crossbar-like architectures, where it could allow massive parallel logic operation. Different RRAM-based IMC logic solutions have been studied so far, from circuit level and architecture to algorithms [3][4][5]. However, RRAM variability is a considerable challenge for these schemes, and only a few have demonstrated experimental results, with limited parallelism [3]. Additionally, all these works use RRAM as a single-bit memory to perform logic operations, giving up to a major feature of most RRAMs, their ability to provide Multi-Level Cell (MLC).

In this work, for the first time, we demonstrate experimentally an RRAM-based IMC logic concept with strong resilience to RRAM variability, a parallel generalization of a concept known as Scouting Logic [6]. Logic operations (NAND, NOR, XOR) are performed on up to 16 parallel RRAM devices (operands), a new milestone for experimental RRAM IMC. Additionally, we show, also for the first time, that this concept can be extended to MLC RRAM and demonstrate an MLC-based IMC 2-bit adder experimentally.

Our technique belongs to the Non-Stateful Logic IMC category, meaning that the results of the operations are obtained after a read operation on multiple cells in parallel that preserves memory endurance [7]. To overcome the RRAM variability issue, we proposed a new smart programming method suitable for binary and Multi-Level Cell solutions.

All experiments have been conducted on a custom $HfO_2$ crossbar 1T1R memory array fabricated in a commercial 130 nm technology node (Fig. 1a). This memory array, incorporating the entire required CMOS periphery, can operate on both Memory Mode (Fig. 1b) and In-Memory Computing Mode (Fig. 1c), thanks to the selection of multiple cells in parallel during a read operation.

## II. NEW SMART PROGRAMMING STRATEGIES

RRAM programming suffers from short-term (relaxation) and long-term (retention) conductance drift, which is a severe limitation for MLC storage and MLC/binary IMC [8]. In this work, we propose a new "smart" programming method to avoid conductance relaxation, called Full-Correction Smart Programming (FC-SP). It is an extension of the more standard algorithm of [8], here called Partial-Correction Smart Programming (PC-SP). Both algorithms use RRAM cycle-to-cycle variability to program the devices in a specific conductance range. The main addition of FC-SP is a wait time Δt (Fig. 2) after the SET operation. This waiting time enables the algorithm to take into account conductance relaxation. Fig. 3 shows a conductance level programmed with PC-SP: the conductance relaxation causes the initial distribution to spread over time, and after one hour, more than 12% of the programmed devices are out of the target conductance range. By contrast, the FC-SP strategy (Fig. 4) results in conductance levels that remain stable, even in the case of MLC storage. For FC-SP, the effect of conductance relaxation is negligible after one hour (less than 1% of cells are out of the target range for the corresponding level 0 in Fig. 4). The impact of different waiting times Δt in the FC-SP algorithm is presented in Fig. 5. Only a minor change is observed between a wait time of 5 s and 30 s. After the waiting time of 5 s (the minimum waiting time allowed by our experimental setup), the FC-SP MLC distributions are stable after one minute (Fig. 6a) and one hour (Fig. 6b). Therefore, in subsequent experiments presented in this paper, the waiting time Δt for FC-SP is always 5 s.

Fig. 7 shows the number of devices out of the target conductance range (Bit Error Count - BEC) through successive iterations. The FC-SP programming strategy needs more iterations to achieve the same programming performance compared to PC-SP. The BEC of the most critical conductance level of an MLC (i.e. the range corresponding to the lower conductance value, level 0 in Fig. 4) rapidly increases during the first 10 s if the cells are programmed with PC-SP due to conductance relaxation, while it remains low for FC-SP (Fig. 8). The retention data of three programmed levels with FC-SP is evaluated over a month (Fig. 9). FC-SP MLC levels are resilient to both short-term relaxation and long-term data retention.

## III. BINARY IN-MEMORY COMPUTING

We designed, fabricated, and tested a crossbar structure (Fig. 1a) that allows the selection of several RRAM devices simultaneously, which is the essential feature to enable IMC. The binary RRAM programmable logic scheme for NAND, NOR, and XOR is shown in Fig. 10. This technique, known as Scouting Logic [6], requires a current reference to differentiate the read combinations and classify the logic output as a "0" or "1". The Scouting Logic concept is here extended to n parallel devices (operands), allowing to perform the n logic operations simultaneously.

First, we investigate the approach using RRAM programmed without any smart programming strategy. Fig. 11 shows experimental measurements of read current distributions of two and four parallel devices read simultaneously (the read voltage $V_{read}$ is always 0.4 V). RRAM conductance variability causes overlap between the read current distributions. Fig 12 shows the corresponding Scouting Logic operation success rate with up to 16 devices in parallel. We see that the resistance variability associated to standard SET/RESET operations makes the Scouting Logic acceptable only for up to four devices in parallel: the overlap between read current distributions causes the success rate to collapse with 8 and 16 devices in parallel. Alongside conductance variability, another drawback of RRAM technology is their degradation during endurance. Therefore, we SET/RESET cycled the RRAM cells one million times and, after each decade of cycling, performed the Scouting Logic operations. The success rate remains stable even after $10^6$ cycles (Fig. 13).

To enhance the Scouting Logic IMC success rate, we then used the FC-SP strategy of section II, in order to reduce the distribution tails and control RRAM resistance intrinsic variability. Fig. 14 highlights the difference between the read current distributions for two and four devices in parallel programmed initially with and without FC-SP. The FC-SP narrows the distributions, making it more suitable for IMC operations. We then show that it is possible to more parallel devices on Scouting Logic IMC mode when FC-SP is adopted (Fig. 15). Fig. 16 shows the difference of performance between Scouting Logic IMC success rate with and without FC-SP. NAND, NOR, and XOR operations with high yield using up to eight devices in parallel and NAND operation using up to 16 devices in parallel are demonstrated with a success rate higher than 98%. FC-SP is, therefore, fundamental to perform IMC operations with a high amount of devices in parallel.

## IV. MULTI-LEVEL CELL IN-MEMORY COMPUTING

We now take advantage of multiple conductance levels in a single device, combined with parallel read, to achieve more complex logic functions in-memory. We demonstrate an RRAM-based Multi-Level In-Memory 2-bit adder. Fig. 17a shows the conventional 2-bits adder with 2 inputs ($A_0A_1$ and $B_0B_1$) based on a half adder and a full adder, with classic Boolean logic gates used to perform this operation (about 48 transistors are necessary to implement this operation using CMOS technology). In [9], a CMOS-based IMC is proposed; however, it requires additional latch circuits to store intermediate data temporarily. Our solution proposes to combine parallel read of two devices storing four conductance levels (equivalent to two bits) to perform the 2-bits adder operation (2 transistors and 2 RRAM, Fig. 17b). The RRAM Based Multi-Level In-Memory 2-bit adder can be implemented naturally in-memory using the crossbar architecture of Fig. 1b following the Table in Fig.17 b. To ease the differentiation of the output logic states, the programmed conductance levels are chosen linearly spaced (Fig. 17b).

Fig. 18 shows the difference of read conductance distributions of the seven possible logic outputs of the proposed Multi-Level In-Memory 2-bit adder. With the initial four conductance levels programmed with the standard PC-SP, the conductance relaxation causes overlaps, especially between the first four logic states, while with FC-SP the overlap effect is minimal. FC-SP reduces overlaps between read conductance levels in parallel and is fundamental to achieve acceptable performance of the proposed Multi-Level In-Memory 2-bit adder. The corresponding error rate is shown in Fig. 19. The error occurs only between adjacent conductance levels. The error rate is high for the PC-SP based levels, but is lower than 5% for the FC-SP.

To go further, we also explored the possibility of reading three devices in parallel in order to implement an In-Memory Multi-Level RRAM based 2 bits adder with 3 inputs ($A_0A_1$, $B_0B_1$, $C_0C_1$). Reading three devices in parallel results in ten possible logic outputs, as shown in Fig. 20. The overlap is excessive for classic computing paradigms, but this strategy involves only soft errors resulting in an approximate sum, enabling to address a wide range of applications where approximate calculation is supported [10].

## V. CONCLUSIONS

This work demonstrates experimentally a smart programming strategy that controls conductance relaxation in Multi-Level Cell programming and stabilizes programmed levels up to more than one month. This strategy proves to be fundamental for both binary and multi-level IMC. We show that binary IMC based on Scouting Logic on an RRAM crossbar is achievable with up to 16 devices (operands) in parallel, a result that had never been obtained experimentally. Moreover, by combining MLC programming strategy and IMC, we demonstrated experimentally, for the first time, an RRAM-based Multi-Level In-Memory 2-bit adder. These results highlight the potential of RRAM IMC logic, and bring this field beyond purely circuit level and architecture studies.

**ACKNOWLEDGMENT:** This work is supported by the ECSEL TEMPO project (826655) and the ANR grant NEURONIC (ANR-18-CE24-0009).

# I – Introduction

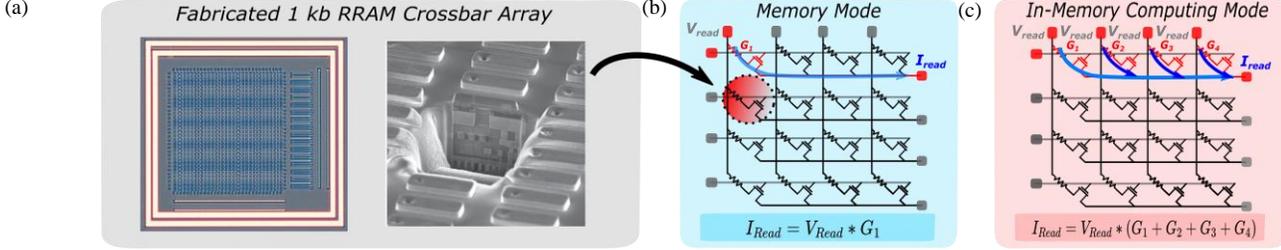

**Fig. 1:** Fabricated 1 kb crossbar RRAM crossbar array based on 1T1R HfO$_2$ stack (a). The proposed architecture can operate on both Memory Mode (b) and In-Memory Computing Mode (c). In Memory Mode, a single device or the full column is addressed and the current is sensed along the row using the current $I_{read}$. On the In-Memory Computing Mode, several devices of the same line are selected simultaneously

# II – New Smart Programming Strategies

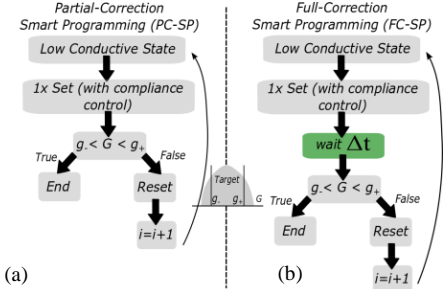

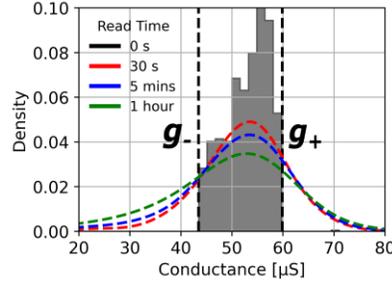

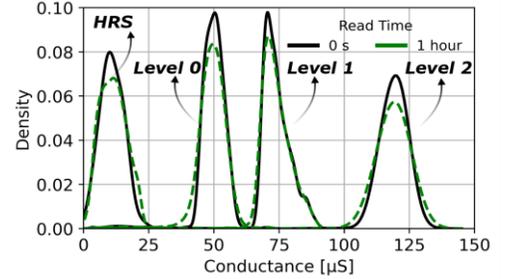

**Fig. 2:** MLC Programming strategies flow: Partial Correction Smart Programming (PC-SP) [8] (a) and Full Correction Smart Programming (FC-SP) (b).

**Fig. 3:** Conductance distribution programmed with PC-SP strategy. The dashed lines represent the relaxation over different read times.

**Fig. 4:** Three conductance levels programmed with FC-SP strategy and the High Resistive State (HRS) just after programming (black) and after 1 hour (green).

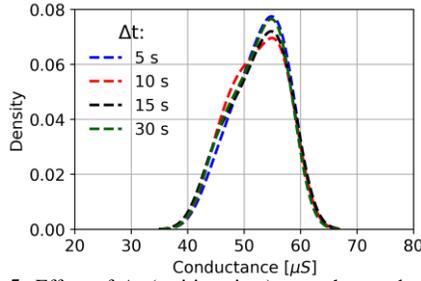

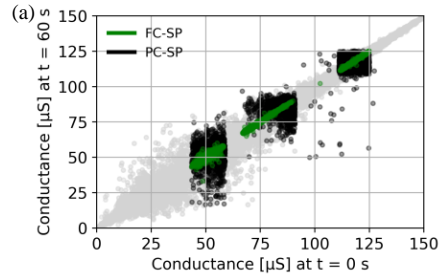

**Fig. 5:** Effect of $\Delta t$ (waiting time) over the conductance distribution of the programmed level 0 after FC-SP.

**Fig. 6:** Relaxation effect for FC-SP and PC-SP after 60 seconds (a) and 1 hour (b). Grey color represents distributions without smart programming. Relaxation occurs on the first seconds. $\Delta t$ is 5 s for FC-SP.

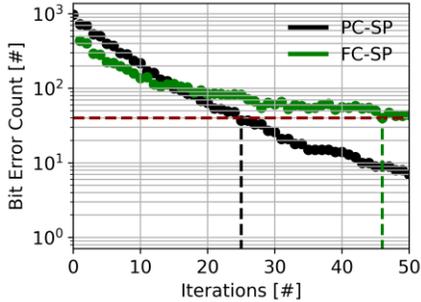

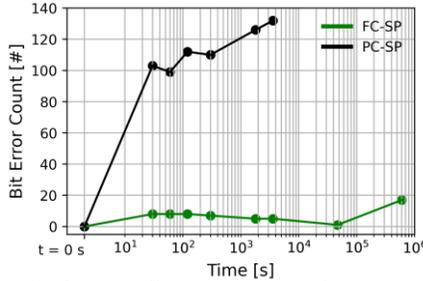

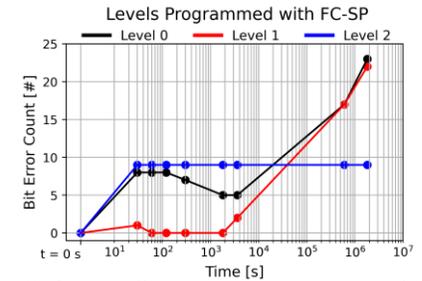

**Fig. 7:** Bit Error Count (BEC) as a function of the number of iterations during PC-SP and FC-SP.

**Fig. 8:** Bit Error Count on time for the most error prone level (level 0 in Fig. 4) programmed with FC-SP and PC-SP.

**Fig. 9:** Bit Error Count on time (up to one month) for the 3 FC-SP programmed levels. HRS is not shown, as there is no error for this state.

# III – Binary In-Memory Computing

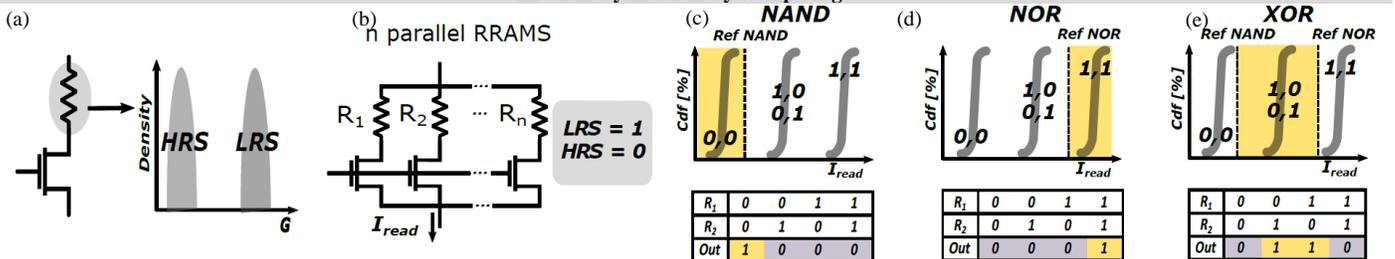

**Fig. 10:** Logic scheme of in-situ logic operations based on the parallel reading of n devices and the resulting conductance distributions used for NAND (a) NOR (b) and XOR (c) operations. The XOR operation is defined as the complementary operation of NAND and NOR.

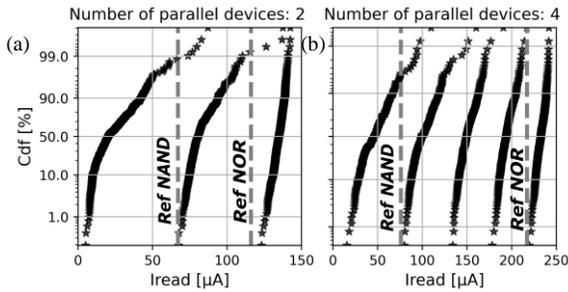
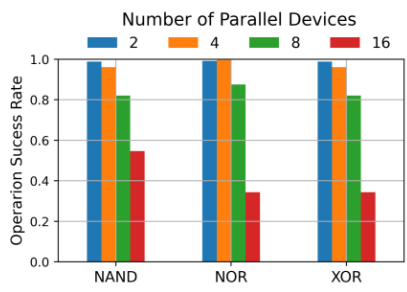
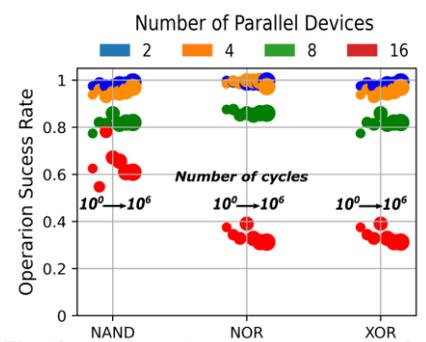

**Fig. 11:** Experimental read current distributions covering all the possible LRS/HRS combinations for two (a) and four (b) parallel devices programmed without smart programming. Optimal current references are shown as dashed lines.

**Fig. 12:** Operation success rate for different number of parallel devices using standard SET/RESET (without smart programming) for NAND, NOR and XOR operations.

**Fig. 13:** Experimental operation success rate after an increasing number of decades of endurance for NAND, NOR and XOR operations.

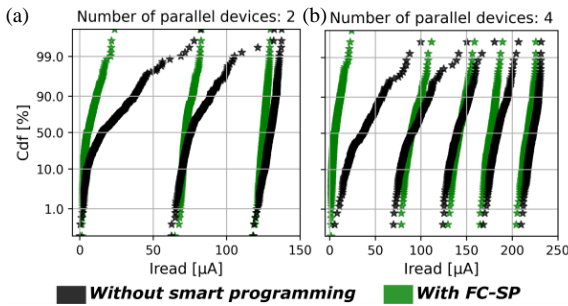
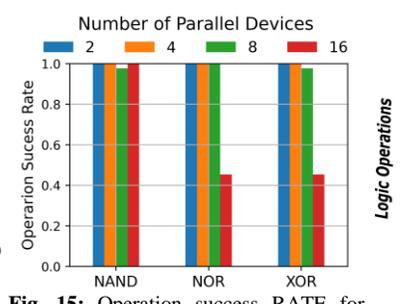
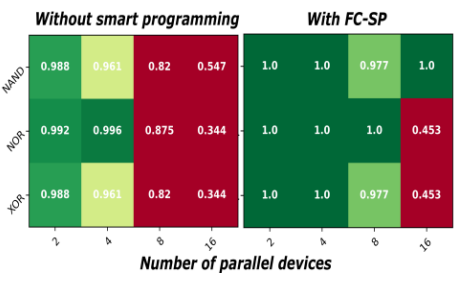

**Fig. 14:** Difference between read current distributions with FC-SP and without smart programming for two parallel devices (a) and four parallel devices (b).

**Fig. 15:** Operation success RATE for different number of parallel devices using FC–SP for NAND, NOR and XOR operations.

**Fig. 16:** NAND, NOR, XOR operations success mapped after programming without smart programming and with FC-SP.

## IV – Multi-Level-Cell In-Memory Computing

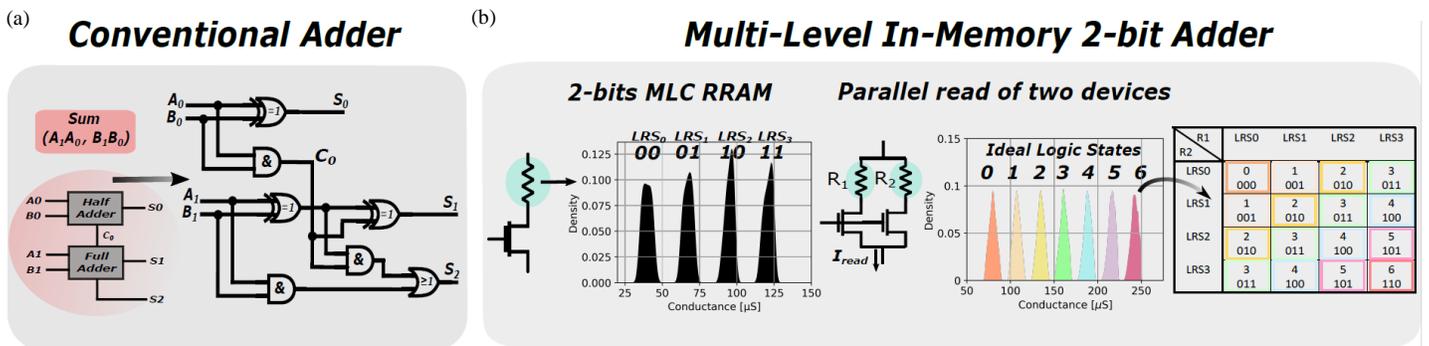

**Fig. 17:** Conventional architecture needed for a 2-bit adder (a) and proposed Multi-Level In-Memory 2-bit adder based on two parallel RRAMs (b).

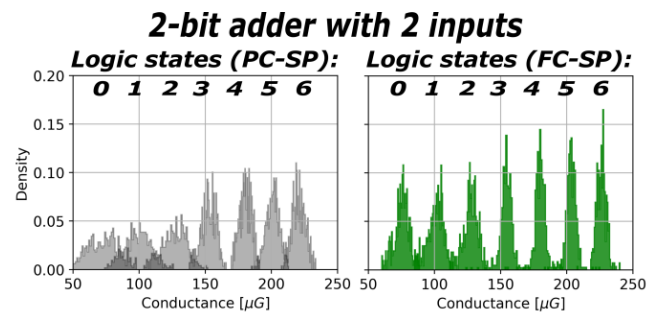
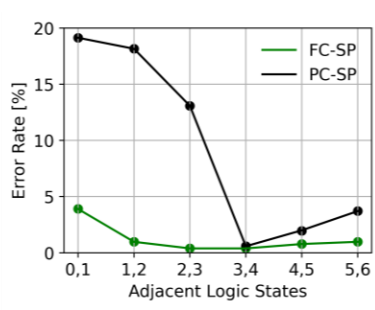
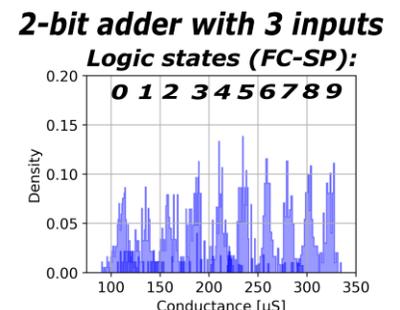

**Fig. 18:** Experimental distributions of the In-Memory 2-bit Adder logic states based on conductance distributions obtained with PC-SP (a) and FC–SP (b).

**Fig. 19:** Error rate between adjacent logic states for the proposed 2-bit adder for PC-SP and FC-SP.

**Fig. 20:** IMC 2-bits adder with 3 inputs (3 parallel devices) suffers from overlap. The adopted MLC strategy is FC-SP.